\documentclass[aps,physrev,superscriptaddress]{revtex4-2}

\usepackage{xcolor}
\usepackage{color}
\usepackage[normalem]{ulem}
\usepackage{float}
\usepackage{gensymb}
\usepackage{graphicx}
\usepackage{subfigure}
\usepackage{dcolumn}
\usepackage{bm}
\usepackage{xcolor}%
\usepackage{amsmath}
\usepackage{amssymb}
\usepackage{verbatim}
\usepackage{mathtools}
\usepackage{epstopdf}
\usepackage{float}
\usepackage{lineno}
\usepackage{listings} %
\usepackage{booktabs}
\usepackage{multirow}
\usepackage{array}

\def\<{\langle}
\def\>{\rangle}

\begin{document}


\title{
At low temperatures, glass-forming liquids relax in a simple way
}

\author{Francesco Rusciano}
\affiliation{Department of Chemical, Materials and Production Engineering, University of Naples Federico II, P.le Tecchio 80, Napoli 80125, Italy.}

\author{Raffaele Pastore}
\affiliation{Department of Chemical, Materials and Production Engineering, University of Naples Federico II, P.le Tecchio 80, Napoli 80125, Italy.}

\author {Francesco Greco}
\affiliation{Department of Chemical, Materials and Production Engineering, University of Naples Federico II, P.le Tecchio 80, Napoli 80125, Italy.}

\author {Walter Kob}
\affiliation{Laboratoire Charles Coulomb, University of Montpellier and CNRS, 34095 Montpellier, France}

\begin{abstract} 
Glass-forming liquids have only a modest tendency to crystallize and hence their dynamics can be studied even below the melting temperature. The relaxation dynamics of most of these liquids shows at a temperature $T_c$, somewhat above the glass-transition temperature $T_g$, a crossover, which indicates the conjunction of two different dynamical regimes. For temperatures slightly above $T_c$, experiments and computer simulations have extensively probed this dynamics on the particle level and identified several universal scaling laws that are often compatible with theoretical predictions. Using large scale computer simulations we extend these studies to temperatures below $T_c$ and find that the relaxation mechanism is qualitatively different from the one found at higher temperatures. We identify new scaling laws that allow to give a simple description of the relaxation dynamics at very low $T$s. 
Specifically we reveal that the cage-escape process is related to rare but large particle displacements that give rise to a distinctive sub-diffusive power-law in the time correlation functions.
This insight helps to advance our understanding on the relaxation dynamics of glass-forming systems at temperatures that are close to the experimental glass transition. 
\end{abstract}
\maketitle

In the last decade impressive advancement in our understanding of glassy systems has been made, progress that was triggered by novel theoretical tools like machine learning, advanced simulation techniques, as well as experimental methods that have allowed to probe the static and dynamic properties of supercooled liquids and glasses in unprecedented details~\cite{schonholtz, parisi_swap, ninarello_swap, kou_granular}. These studies have permitted to reach, on the particle level, a satisfying understanding of the relaxation dynamics of glassy systems for temperatures that are around or above the critical temperature $T_c$ of Mode-Coupling Theory (MCT)~\cite{gotze_book}. Much less is known, however, about the dynamics below $T_c$, since at these temperatures the relaxation times are usually by many orders of magnitude larger than the vibrational ones, making it very challenging to examine the details of the $\alpha$-relaxation. (Although network glass-formers can be simulated also below $T_c$~\cite{horbach}, their relaxation dynamics is very different from the one of fragile glass-formers~\cite{garrahan2003coarse, vogel2004temperature, coslovichnetwork}.) Also from the theoretical side the sub-$T_c$ regime is not studied well, since it is difficult to make analytical predictions that go beyond the ones made by the simplest version of MCT~\cite{fuchs,szamel, janssen}. Although there is consensus that close to and below $T_c$ the divergence of the relaxation time as predicted by MCT must be cut off, the details of the process governing the $\alpha$-relaxation, often called ``hopping process''~\cite{charbonneau2014hopping,schweizer2003entropic,sastry1998signatures}, have not been elucidated so far. The goal of the present work is thus to investigate the relaxation dynamics in this temperature regime on the level of the particles. We find that around and below $T_c$ this dynamics displays universal aspects that make it surprisingly simple, a result that will allow to advance our theoretical understanding of deeply supercooled liquids. \\[1mm]

\textbf{System}\\ 
Our system is a binary Lennard-Jones mixture that has been often used in the past to study the relaxation dynamics of simple glass-formers~\cite{kob1994scaling}. We use a slight modification of the interaction parameters which suppresses crystallization~\cite{schroder}, but does not affect the (above $T_c$) relaxation dynamics, thus the critical temperature $T_c$ remains around $0.435$~\cite{kob1994scaling}. The number of particles is 32.400 and the simulations were done at constant volume, with $T=0.375 \approx 0.86 T_c$ being the lowest investigated temperature. At the lowest $T$s, the length of the runs was $2\cdot 10^9$ time-steps, and we averaged over 4 independent samples.
More details on the simulations are given in the SI. In the following we discuss the dynamics of the larger particles since they are in the majority (80\%). \\[1mm]

\begin{figure}[ht]
\includegraphics[width=0.7\linewidth]{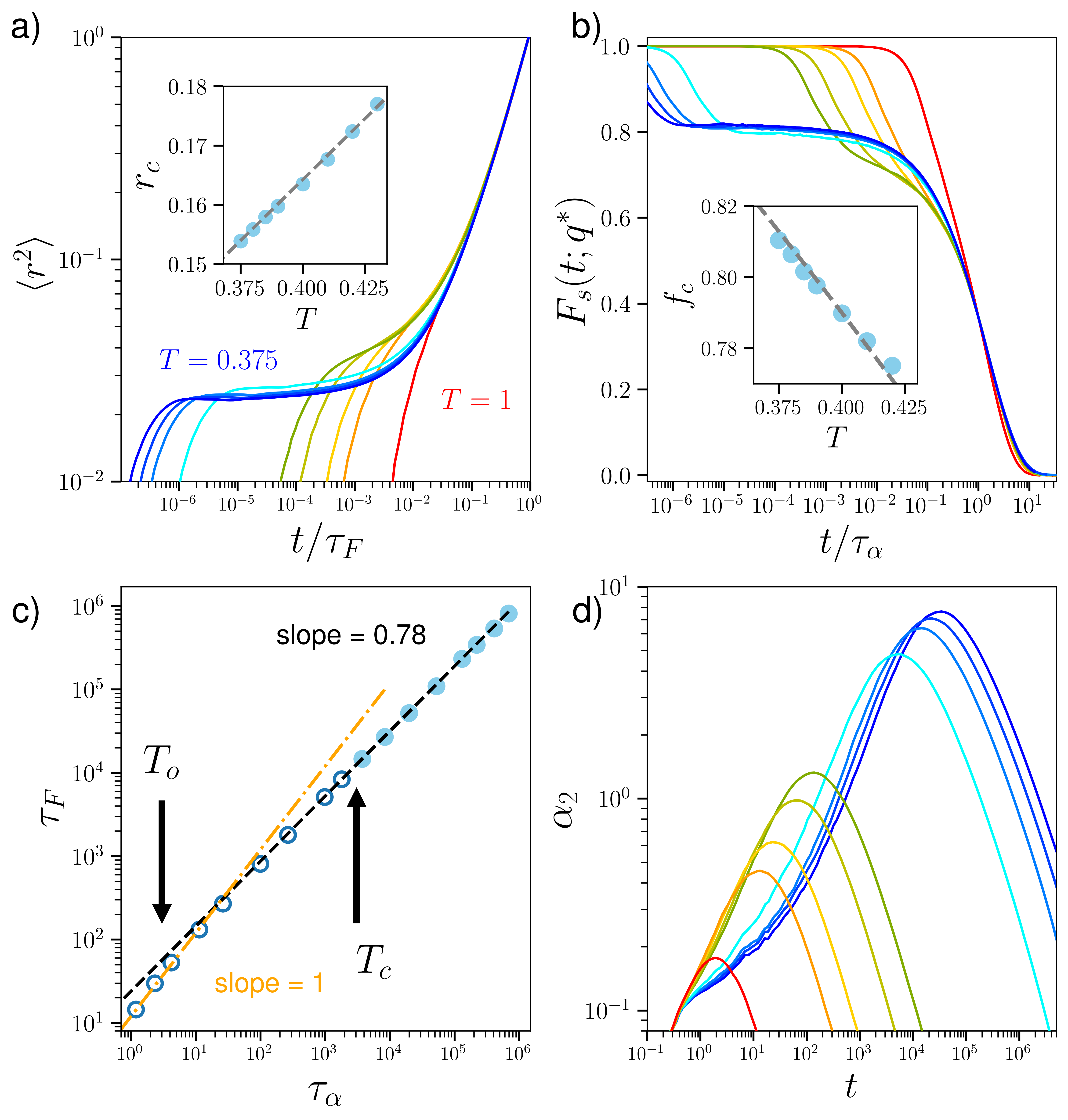}
\caption{
(a) MSD as a function of $ t / \tau_F$, for the temperatures $T=1.0,$ 0.6, 0.55, 0.5, 0.475, 0.4, 0.385, 0.38, and 0.375, from right to left. Inset: MSD plateau height $r_c$ as a function of temperature for $T<T_c$. The dashed line is a linear fit. (b) The self intermediate scattering function (SISF) computed at wave-vector $ q^*=7.25$, the location of the main peak in the static structure factor, as a function of $ t / \tau_\alpha$,  for the same temperatures as in (a). Inset: Plateau height of the SISF as a function of temperature for $T<T_c$. The dashed line is Gaussian fit $f_c \propto e^{-a T^2}$. 
(c)  $\tau_F$ versus $\tau_\alpha$. The orange line is a linear fit to the data for $T \geq T_o$, $T_o$ being the estimated onset temperature of glassy dynamics for this system, $T_o\approx 0.8$ (\cite{gpucoslovich, nandi_2021});  the black line is a power-law fit $\tau_F \propto \tau_\alpha^{0.78}$ to the data for $T<T_c$ (filled symbols).  (d) Non-Gaussian parameter as a function of time for the same temperatures as in panel (a). 
}
\label{fig:fig1}
\end{figure}

\textbf{Results}\\
Figure~\ref{fig:fig1}a presents $\langle r^2(t)\rangle$, the Mean Squared Displacement (MSD) of a particle, as a function of the rescaled time $t/\tau_F$, where the Fickian time scale $\tau_F$~\cite{rusciano2022fickian} is defined as $\tau_F=\sigma^2 / 6D$, with $D$ the diffusion constant obtained from the MSD at long times. 
By construction, this rescaling makes that in the diffusive regime all MSD's fall on a master curve. For $T$'s slightly above $T_c$, data follow a master curve also in the time range in which the particles start to leave the plateau, i.e., at the end of the caging regime. This behavior is compatible with MCT which predicts a $t$-dependence of the form $r_{c}^2+a(t/\tau_F)^{b}+c(t/\tau_F)$, with cage size $r_{c}$ and exponent $b$, both independent from $T$. 
Previous fitting results~\cite{kob1995testing1} indicated $r_c \approx 0.2$ and $b\approx 0.5$, fully compatible with the present results.
Interestingly, one finds that for temperatures below $T_c$ the MSD's no longer fall onto a master curve, indicating that the escape of the particles from the cage is related to a microscopic mechanism that differs from the one at higher $T$'s. The height of the plateau, i.e., the size of the cage, now depends on $T$, see Inset of panel (a). This dependence is linear in $T$, at variance with the $T^{1/2}$-dependence expected for a harmonic system. Hence one concludes that, despite the fact that the particles are trapped for a significant amount of time, the cage has a significant anharmonic shape~\cite{sastry1998signatures}, in agreement with previous studies on the configurational entropy~\cite{ozawa} and with the result that $T_c$ corresponds to a transition in the dynamic exploration of the potential energy landscape~\cite{sciortino,cavagna}.  
Panel (b) shows that also the Self-Intermediate Scattering Function $F_s(q,t)$ (SISF) plotted versus the $\alpha$-relaxation time $\tau_\alpha$, displays the same qualitative scenario emerging from panel (a). In agreement with MCT, the curves at intermediate $T$s fall onto a master curve with a decay from a plateau that is well described 
by a power-law $f_c- B(T)t^b$, with the non-ergodicity parameter $f_c$ and the von Schweidler exponent $b$, both independent of $T$. For low $T$s, the curves no longer fall on a master curve with $f_c$ increasing with decreasing $T$, see Inset of panel 
(b). Thus we find again qualitative differences in the dynamics at intermediate and deeply supercooled conditions (approximately above and below $T_c$).
Panel (c) shows the relation between the two timescales $\tau_\alpha$ and $\tau_F$. For high and intermediate temperatures one finds the expected Stokes-Einstein relation, i.e., $\tau_F \propto \tau_\alpha$. About 10\%  above $T_c$ ($T\approx 0.475$), this dependence is replaced by a fractional Stokes-Einstein relation $\tau_F \propto \tau_\alpha^c$~\cite{sengupta2013breakdown,das2022crossover}, with an exponent $c$ around 0.78, indication of strong dynamical heterogeneities (DH) in the sample~\cite{ediger_review,richert_review}. 
One indicator for the strength of these DH is the non-Gaussian parameter $\alpha_2(t)=5\langle r^4(t)\rangle/3\langle r^2(t) \rangle^2 -1$, the $t$-dependence of which is presented in panel (d). One recognizes that at high and intermediate temperatures the rise of $\alpha_2(t)$ is essentially independent of $T$ and is well described by a power-law, in agreement with earlier studies~\cite{nandi_2021}. For temperatures below $T_c$, this is no longer the case, showing that there is a change in the nature of the $T$-dependence of the DH. The $T$-dependence of $\tau^*$, the time at which $\alpha_2$ reaches its maximum, is presented in Fig.~\ref{figSM_tauF_taustar}. 

Overall, Figure~\ref{fig:fig1} demonstrates that, upon changing $T$, there are two different dynamic regimes. We do not claim, however, that $T_c$ is the exact boundary between these two different dynamical behaviors. Rather, this temperature should be regarded as indicative of a point at which where the nature of the dynamics changes, and we emphasize that we reach this conclusion from the analysis of the raw data only, without making any fit inspired by MCT.

\begin{figure}
\includegraphics[width=1\linewidth]{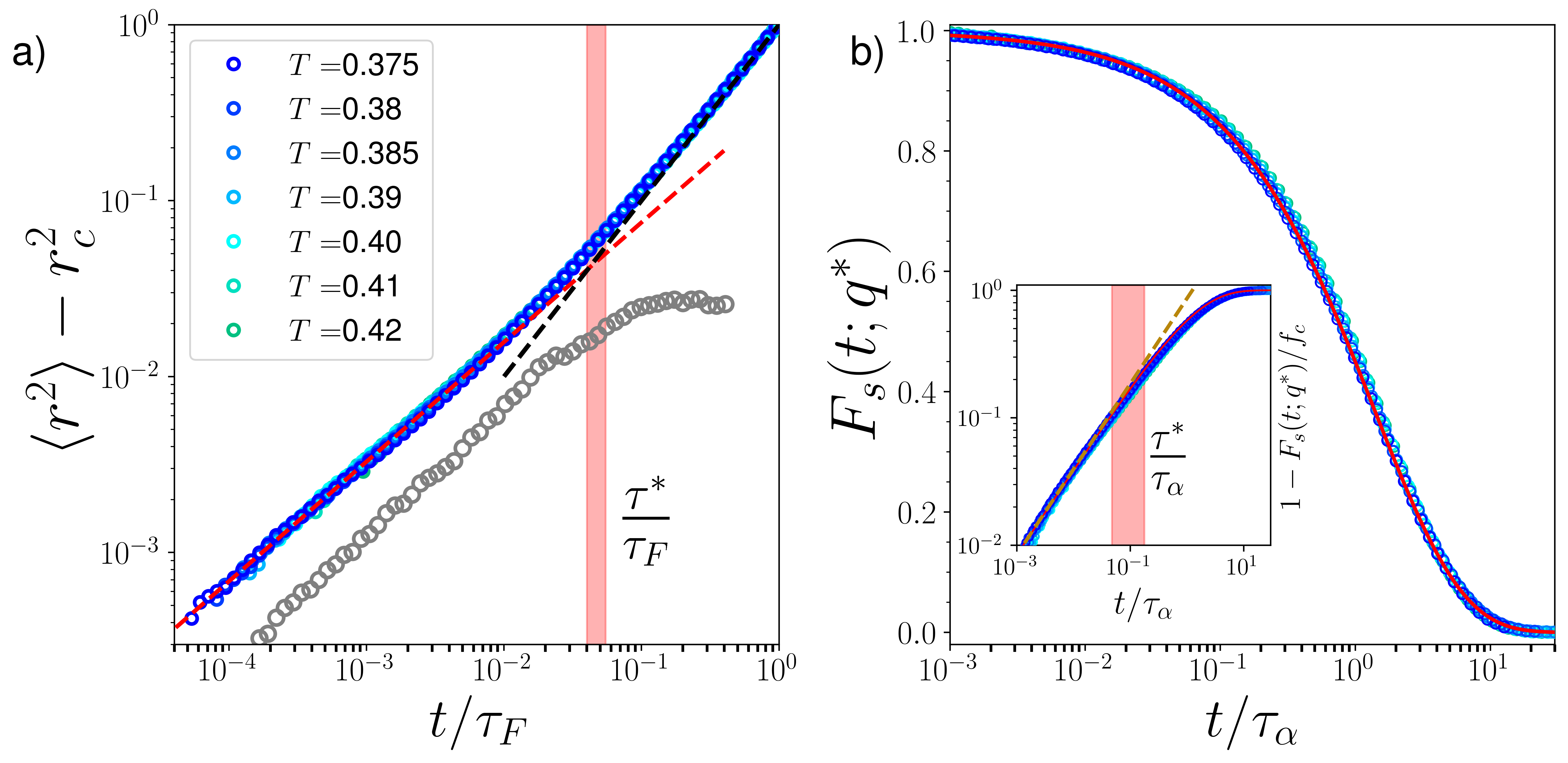}
\caption{
(a) MSD - $r_c^2(T)$ as a function of $ t / \tau_F$, for different temperatures in the sub-$T_c$ regime. Black and red lines are $t$ and $t^{2/3}$ power-laws, respectively. Contribution MSD$_{tail}$ (see text) for $T=0.375$ is plotted as gray points.
(b) SISFs rescaled by the non-ergodicity factor $f_c(q^*)$ as a function of $ t / \tau_\alpha$, for the same temperatures as in (a). Red dashed line is a fit with a stretched exponential with exponent $\beta=2/3$. 
Inset:  Red dashed line indicates the short time expansion $\propto t^\beta$ of the stretched exponential fit.
The red shadowed regions in the panels indicate the times $\tau^*$ (see Fig.~\ref{figSM_tauF_taustar}) at the different displayed temperatures. 
}
\label{fig:fig2}
\end{figure}

In the following we focus on the dynamics at low $T$s and demonstrate that it shows system-universal features that permit to describe it in a simple manner. Figure~\ref{fig:fig2}(a) presents $\langle r^2(t; T)\rangle-r^2_c(T)$ as a function of $t/\tau_F$. One recognizes that the time-dependence of this difference is independent of $T$ over several decades in time, i.e., the escape from the cage is independent of $T$. For times shorter than $\tau^*$ (vertical bar) the increase of $\langle r^2\rangle-r_c^2$ is a power-law with exponent $b'=0.67\pm 0.02 \approx 2/3$;
for times larger than $\tau^*$, the curve starts to bend upwards to enter the Fickian regime around $\tau_F$ ~\cite{rusciano2022fickian, rusciano2023}.
To the best of our knowledge, the presence of such a unique sub-diffusive behavior after the caging regime, at temperatures significantly below $T_c$, has never been observed in earlier studies of supercooled liquids.
The emergence of a low-$T$ dynamics that, up to a scaling factor, is independent of $T$ can be observed also in $F_s(q,t)$.  In Fig.~\ref{fig:fig2}(b) we present the time dependence of $F_s(q,t;T)/f_c(T)$ and one recognizes that all curves below $T_c$ fall indeed on a master curve when plotted as a function of $t/\tau_\alpha$. 
Note that this master curve is observed in the whole $\alpha$-regime as well as in the late $\beta$-regime (see Inset), i.e., throughout a time-range that is much larger than the one found for temperatures above $T_c$~\cite{kob1995testing2}. 
The master curve is very well described by a stretched exponential, $\exp\{-(t/\tau_\alpha)^\beta\}$, with $\beta=0.67 \pm 0.02 \approx 2/3$, i.e., the exponent is the same as the one we found for the MSD in panel~(a). To the best of our knowledge, the identification of such well defined sub-diffusive post-caging scalings, with $b'=\beta$, has never been reported before for glass-forming liquids. This demonstrates that at low temperatures the $\beta$-and $\alpha$-processes for the MSD and $F_s(q,t)$ are strongly related to each other, in contrast to the case at higher temperatures for which the von Schweidler exponent (independent of $T$) and the stretching exponent (often varying with $T$~\cite{sen,das2022crossover,sastry1998signatures}) are different~\cite{kob1994scaling}.
For the sake of completeness we show in Fig.~\ref{figSM_beta}(a) the temperature-dependence of $\beta$ as obtained from our data, throughout a range of temperatures, $5.0 \geq T \geq 0.375$.
It is indeed evident that at high and intermediate $T$s a significant $T-$dependence of $\beta$ exists, but it saturates at very low $T$, indicating that the asymptotic regime has been reached.

\begin{figure*}
\includegraphics[width=1.0\linewidth]{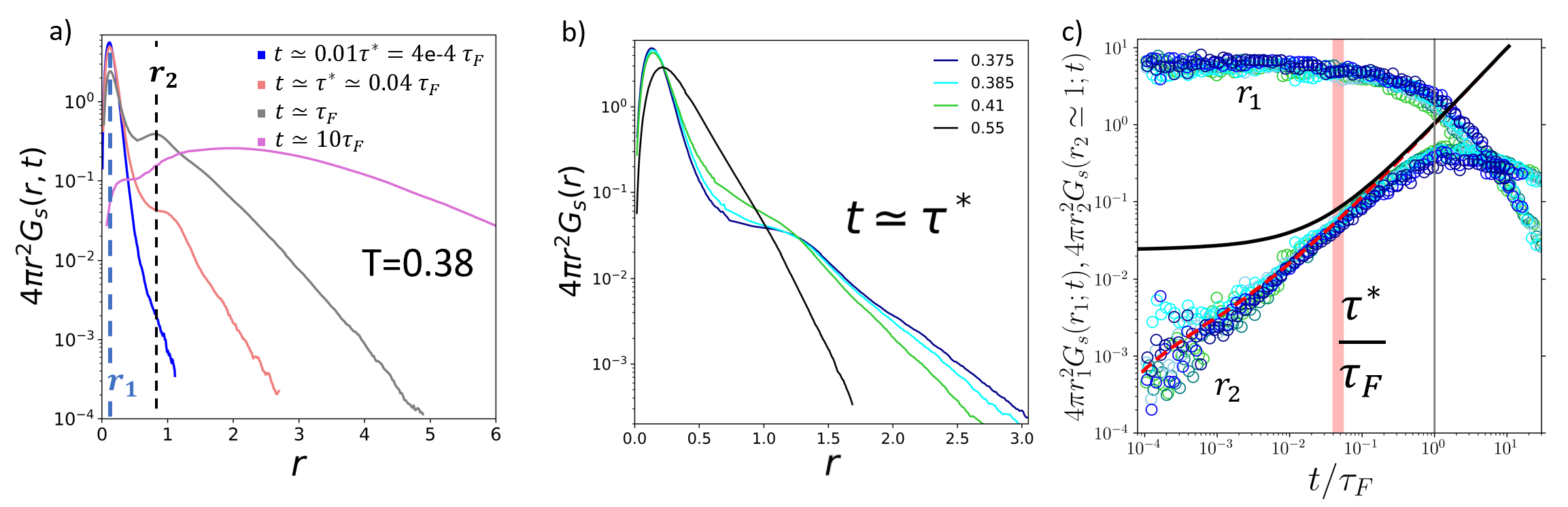}
\caption{
(a) $r$-dependence of $4\pi r^2 G_s(r,t)$ at $T=0.38$ for different times, as indicated in legend.
(b) $4\pi r^2 G_s(r,t)$ as a function of $r$, for different temperatures $T$, indicated in the legend, at $t=\tau^*(T)  $. The vertical dashed lines indicate the length scales $r_1$ and $r_2$, the size of the cage and the hopping distance, respectively.
(c) Value of $4\pi r^2 G_s(r,t)$ in $r=r_1(t)$ and $r\simeq0.9$, as a function of rescaled time. $\< r^2( t)\>$$-r_c^2$ is  plotted as red dashed line, $\< r^2( t)\>$ is plotted as black solid line, for $T=0.375$.  Red shaded regions in figure indicate the range where the $\tau^*$ (see Fig.~\ref{figSM_tauF_taustar}) are located at the different displayed temperatures.
}

\label{fig:fig3}
\end{figure*}

Deeper insight into the nature of the observed low-$T$ dynamics after the caging regime is obtained by probing the self Van Hove function $G_s(r,t)$, i.e., the probability that a particle has moved in time $t$ by a distance $r$. Figure~\ref{fig:fig3}a shows that $4 \pi r^2G_s(r;t)$ peaks at a distance $r_1$ at short times, i.e., it shows a Gaussian core due to the particles that rattle inside the cage they were in at $t=0$; the peak position remains almost fixed at $r_1\approx r_c$ for a long time (up to around $\tau_F$, see below).
For times around $\tau^*$ the distribution shows a shoulder which evolves into a second peak at a distance $r_2\approx 0.9$, i.e., close to the position of the main peak in the radial distribution function. This second peak is due to the particles that have left their original cage and have effectively hopped to the location that corresponds to the first neighbor distance~\cite{sastry1998signatures}. This second maximum becomes predominant at long times ($t>\tau_F$) and subsequently starts to move to larger distances, while the first peak decreases and eventually fades away, indicating that the dynamics has become diffusive.
Remarkably, for all times the distribution displays an exponential-like tail, present even at very short times ($t=0.01\tau^*$), i.e., the original cages start to ``leak'' very early, and this leads to the MSD detaching from the plateau. This exponential-like tail persists even at very long times ($t>\tau_F)$, where Fickian non-Gaussian Diffusion~\cite{rusciano2022fickian,rusciano2023universal,miotto2021length} is at play ($\alpha_2$ has not yet decayed to zero).
For temperatures above $T_c$, Chauduri {\it et al.} have found that exponential tails are present in a variety of glass-formers~\cite{chaudhuri2007universal}, and thus our finding generalizes these results to the $T$-range below $T_c$. The observation that around $\tau^*$ the exponential tail starts at a distance $r \approx 1$, indicates that at low temperatures the effective hopping distance is given by the nearest neighbor distance, in contrast to the finding of Ref.~\cite{chaudhuri2007universal}, for which a roughly three times smaller distance was found for $T> T_c$, a value that is compatible with the conclusions reached by the direct analysis of the cage-jumps within single particle trajectories~\cite{pastore2015dynamic}. Thus, this is further evidence that the dynamics above $T_c$ differs qualitatively from the one below $T_c$.  
Figure~\ref{fig:fig3}b illustrates the qualitative differences in the shape of $G_s(r,t)$ for temperatures above/below $T_c$ at fixed $t=\tau^*(T)$. At intermediate temperatures, $T=0.55$, the distribution shows a single maximum and no sign of a shoulder is visible, i.e., the caging core is broad and connects smoothly to the exponential tail. As temperature is lowered significantly below $T_c$, the width of this core shrinks, and a pronounced shoulder is present before the exponential tails. We note that at low $T$$G_s(r,t)$ at $t=\tau^*$ seems to converge to a $T$-independent master-curve,  hinting that the beginning of the $\alpha$-relaxation process tends to become independent from $T$ (if measured on a rescaled time scale) and hence ``simple", coherent with the findings of Figs.~\ref{fig:fig1} and~\ref{fig:fig2}.

Figures~\ref{fig:fig3}a and b also demonstrate that below $T_c$ the dynamics is much more heterogeneous than above $T_c$ in that,  
at the lowest temperatures and up to a time of order $\tau_F$, we have markedly separate populations of particles: Those that are still caged and those that have escaped far from their original cages. This is in contrast to the case of higher temperatures for which this difference is significantly less pronounced.
For intermediate $T$s, the escape far from the cage is due to a sequence of rare 
(short) jumps of length around 0.3~\cite{chaudhuri2007universal}. Instead, at very low temperatures, we observe a few rare events in which particles hop 
by a distance of $O(1)$, before vibrating again in a new cage at $r_2$.  (For the lowest $T$ one indeed sees in Fig.~\ref{fig:fig3}b a further weak peak at $r\approx 2$.) 
In both cases, exponential tails are the result of successions of rare events. %
The relative importance of the two peaks is quantified in Fig.~\ref{fig:fig3}c where we plot the $t$-dependence of their height for the lowest temperatures. 
The peak at $r=r_1$ is the only one present at really short times, and its magnitude remains basically constant for $t\lesssim \tau^*$, i.e. in the time-window
in which one observes the 2/3 power-law growth in the MSD discussed above. In this time-range, no secondary maximum exists, but we still can read off the value of $G_s(r,t)$ at $r_2=0.9$, and one finds that it grows with a 2/3 power-law (see red dashed line). 
For larger times, $\tau^* \leq t \leq \tau_F$, the shoulder at $r_2$ becomes a real peak and reaches its maximum height, before decreasing at very long times. Only for times well beyond the Fickian onset  
this secondary peak becomes the highest one, which demonstrates that Fickian diffusion does not imply that all particles have left their original cage.
In conclusion, the sub-diffusive growth of the MSD is due to particles that have hopped once (secondary maximum) and those anomalously far from their original cages (exponential tails).
The fact that in this low-$T$ regime this simple short-time dynamics depends only on $t/\tau_F$ is highly non-trivial, indicating a \textit{distinctive uncaging process} of the particles. 

\begin{figure}
\includegraphics[width=0.7\linewidth]{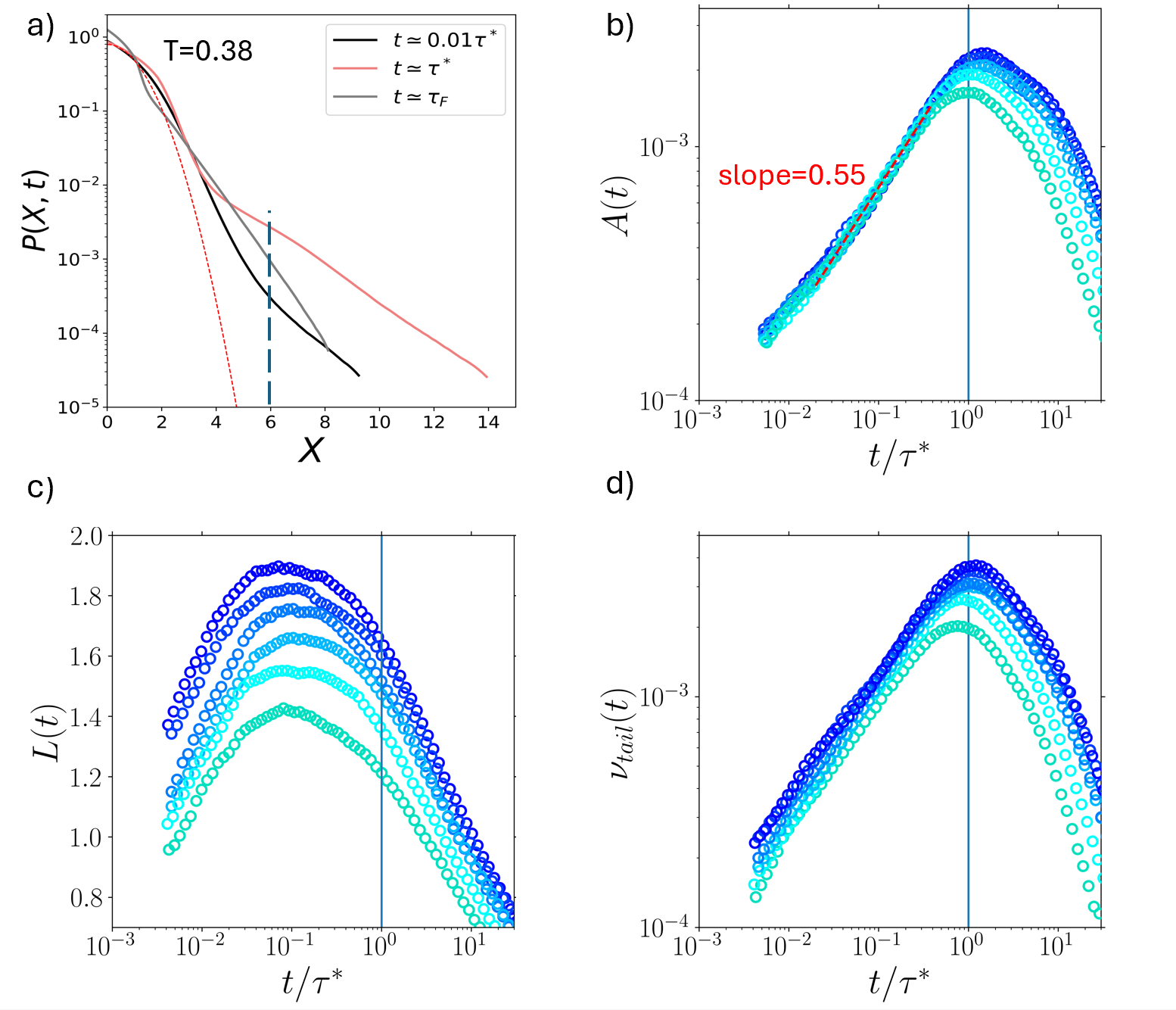}
\caption{
(a) Normalized 1D displacement distribution function as a function of scaled distance $X$ for different times at $T=0.38$. The vertical dashed line is located at $X=X_0$. The red dashed line is a normal Gaussian. 
(b) and (c) show the parameters of the exponential fit to the tails, as a function of $t/\tau^*$ for the lowest temperatures. 
(d) $\nu_{\rm tail}(t)$ as a function of $t/\tau^*$. 
Colors as in Fig.~\ref{fig:fig2}.\\
}
\label{fig:fig4}
\end{figure}

To clarify further the link between the $2/3$ power-law dependence documented in Fig.~\ref{fig:fig2} and the exponential tails in $G_s(r,t)$, we present in Fig.~\ref{fig:fig4}a the one-dimensional Van Hove function at the lowest $T$s. 
To compare the distribution at different times, we plot this probability $P(X,t)$ as a function of the reduced variable $X= { x}/{\sqrt{\< x^2( t)\>}}$. For a Gaussian process, this rescaling results thus in a universal time-independent master-curve: $G( X)=\sqrt{{2}/{\pi}} \exp[-X^2/2]$, included in the figure as well.
For large $X$ all distributions shown in panel (a) display an excess probability with respect to $G(X)$. We note that in this representation, the double-peak seen in Fig.~\ref{fig:fig3}(a) is not visible anymore, but is replaced by a marked shoulder for times around $\tau^*$, where deviations from $G(X)$ are most pronounced.
For large displacements, i.e., $X \geq6=X_0$, which is beyond the shoulder at $r_2$, tails at all times are well described  by an exponential $A(t)\exp[-(X-X_0)/L(t)]$. [Note that, at $t=\tau^*$, the condition $X>X_0$, corresponds in the $3$D real-space to the condition $r>r_2$, since $6 \langle r^2 (\tau^*) \rangle \approx 0.9$.] One sees immediately that the decay length of the tail, $L(t)$, does not change much as a function of time, while the amplitude $A(t)$ does.
Panels (b) and (c) present, respectively, the time evolution of the amplitude $A(t)$ and of the length scale $L(t)$. $A(t)$ increases significantly (by more than one order of magnitude) for $t<\tau^*$ and peaks at $t\simeq \tau^*$. Remarkably, the curves for the different $T$'s fall onto the same master curve before $\tau^*$, showing a power-law with an exponent $\simeq 0.55$. We note that for temperatures above $T_c$ no such master curve is observed (see Fig.~\ref{figSM_A}). 
Also two other features of $A(t)$ for $t> \tau^*$, namely their similar $t$-decay and the presence of a shoulder (at around  $t=8\tau^*$) are not found at higher temperatures (see Fig.~\ref{figSM_A}). These observations show that we are here probing a $T-$regime in which the dynamics is qualitatively different from the one at/above $T_c$, and becomes independent of $T$, if time is appropriately rescaled.
Also $L(t)$ shows non monotonic behavior, panel (c), with a $t-$dependence that is similar for all temperatures, with a maximum approximately at $t=0.1\tau^*$. Throughout the examined $t$-range, however, $L(t)$ at fixed $T$ varies at most by $35\%$ of its maximum value. Thus one concludes that it is the amplitude of the tails, rather than their extent, which is the key ingredient for the DH. (In Fig.~\ref{figSM_l} we report the time-evolution of the dimensional decay length $l(t)=L(t) \sqrt{\< x^2( t)\>}$ and one sees that the variation with time is not very pronounced.)
In Fig.~\ref{fig:fig4}d we present $\nu_{\rm tail}(t)$, the fraction of particles that have made an $X$-displacements larger than $X_0$ (computed as the total area $A(t)L(t)$ under the tail of the distribution function). We recognize that up to times $\tau^*(T)$ this fraction shows a nice power-law behavior, with an exponent and prefactor that is independent of $T$. Hence, this is further evidence that at these low temperatures the relaxation mechanism is simple, i.e.,  independent of temperature.

Finally we make a quantitative comparison between the MSD given by the particles in the tail of the distribution and the total one. 
In the Supporting Information we give an analytic formula for the ratio $\< r^2( t)\>_{\rm tail}/\< r^2( t)\>$, where $\< r^2( t)\>_{\rm tail}$ is the just mentioned contribution to the MSD. It is readily shown that, in leading order, this ratio is $\propto A(t)L(t) \hspace{5pt} (= \nu_{\rm tail}(t)$).
The complete time-dependence of $\< r^2( t)\>_{\rm tail}$ is included in Fig.~\ref{fig:fig2}a as well (for the sake of clarity only for $T=0.375$). 
The figure shows that $\< r^2( t)\>-r_c^2$ and  $\< r^2( t)\>_{\rm tail}$ have the same $t^{2/3}$ time-dependence at short times (below $\tau^*)$, and are in fact close to each other. The same holds for the other temperatures below $T_c$. Hence we conclude that the scaling at the beginning of the $\alpha$-relaxation can be understood quantitatively from the exponential tail in the Van Hove function, i.e.,  rare escaping events account for the sub-diffusive $t^{2/3}$-departure from the plateau in the post-caging regime.
\\

\textbf{Discussion and Outlook:}

Our exploration of the relaxation dynamics at temperatures well below $T_c$ reveals that this dynamics shows unexpected features, some of which render the description of the dynamics more complex, while others simplify it.

The fact that the size of the cage decreases linearly with $T$, i.e., that the cage is strongly anharmonic~\cite{sastry1998signatures}, is surprising since the MSD shows that particles are trapped in their cage for times spanning several decades of their oscillation period, indicating that the cage
is very stable. The linear $T$-dependence of $r_c$ indicates that, at $T$s somewhat below $T_c$, the shape of the cage is not harmonic but has significant higher order components, a result that is important for quantities like the specific heat~\cite{horbach1999specific, ozawa2018ideal}. The marked presence of these anharmonic effects also implies that the prediction of MCT that at $T_c$ the size of the plateau shows a square-root singularity~\cite{gotze_book} is not correct for the present system, and we deem that this discrepancy may not be a particularity of the glass-former we studied, but is instead a quite general result.

The main result of this work is the finding that around $T_c$ the relaxation dynamics changes qualitatively, an insight that has been expected on general arguments~\cite{charbonneau2014hopping,schweizer2003entropic,sastry1998signatures,bouchaud1987anomalous,porpora2022comparing}, but never been probed and documented in detail. While previous studies have shown that the dynamics above $T_c$ is described well by MCT~\cite{kob1994scaling, kob1995testing1, kob1995testing2}, we show here that the one below this temperature shows features that are absent at intermediate and high $T$s. In particular, we find that at low $T$ the process that allows the particles to leave their cages is independent of temperature, if time is measured in terms of the Fickian time $\tau_F$. The Van Hove function shows that this process is closely related to rare events in which the particles hop by a distance that is close to the nearest neighbor distance, in contrast with the case of higher $T$s, for which the jump length is significantly smaller. In conclusion, rare cage-escaping events give rise to a power-law dependence of the MSD below $\tau^*$, i.e., short (rescaled) times, with an exponent that is identical to the stretching exponent found in the $\alpha$-relaxation process, thus at long times. This equality shows that at low $T$s the $\beta$-process is linked to the $\alpha$-process in a way that is much more direct than at $T$s above $T_c$. This implies that the rich and complex relaxation dynamics predicted by MCT above $T_c$ is replaced by a significantly simpler escape problem.

The fact that, at low $T$, one can clearly identify a hopping length that is given by the nearest neighbor distance, indicates that the relaxation process is directly related to simple structural length scale of the liquid. This makes one to expect that the results presented here are generic, i.e., do not depend on the details of the glass-former considered. 
Indeed, one can expect that, once the temperature is significantly below $T_c$, the relaxation dynamics will distinctly involve a simple activated process~\cite{de2017microscopic} with an activation energy that depends on the strength of the local bonding, i.e., a short-range property of the system. 
This conclusion is supported by the experimental observation that the temperature-dependence of the $\alpha$-relaxation time is independent of the glass-former for temperatures below $T_c$, once the fragility of the glass-former has been scaled out~\cite{hess1996parametrization,rossler1998universal,rossler1996}. On the other hand, close to $T_c$, the complex non-linear dynamics as described by MCT makes the $T$-dependence of the relaxation time to depend on many details of the interactions, thus determining the fragility of the glass-former. 

The above conclusions are based on the analysis of typical liquid state observables, like $G_s(r,t)$. It is, however, also interesting to characterize the relaxation dynamics below $T_c$ by probing directly the properties of the particle trajectories and identifying the jump processes via a temporal filtering of these trajectories~\cite{pastore2014cage,pastore2015dynamic,jiang2020efficient,helf2014}. Such an analysis will allow to study the presence of memory effects in the dynamics and hence to establish a connection with the visco-elastic behavior of the system at low $T$. 

Furthermore we note that the quantities presented here are all one-particle observables. As a complementary perspective, it is therefore of interest and important to probe how collective quantities, such as the intermediate scattering function $F(q,t)$ and correlation functions allowing to monitor dynamic heterogeneity, depend on temperature, i.e., whether also they show a simple $T-$dependence below $T_c$. From the experimental results mentioned above, Refs~\cite{hess1996parametrization,rossler1998universal}, one expects that this is indeed true, but a broader verification on the level of the particles is certainly necessary. Also, it is important to study to what extent the results presented here are generic,
i.e., whether the here presented simple relaxation dynamics at temperatures below $T_c$ is a common feature of glass-forming liquids.
Present day computer simulations do allow to investigate this question, and hence such studies should be done since they will allow to verify whether the relaxation dynamics of deeply supercooled glass-formers is actually simpler than the one close to $T_c$. If this is indeed the case, this will be a major step forward in our understanding of glassy dynamics. 

\vspace*{10mm}

{\bf Acknowledgments:}  We acknowledge financial support from the projects MUR-PRIN 2022ETXBEY, ``Fickian non-Gaussian diffusion in static and dynamic environments" and MUR-PRIN 2022 PNRR P2022KA5ZZ,  funded by the European Union – Next Generation EU. W.K.~is a senior member of the Insitut Universitaire de France.

{\color{brown}{

}}

\newpage
\clearpage
{\bf Methods and Materials}\\
{\it Details on the simulations:} We consider a binary Lennard-Jones mixture with particles that we denote by $A$ (80\%) and $B$ (20\%). The interaction between particles of type $\alpha$ and $\beta$ $\in \{A,B\}$ is given by

\begin{equation}
    V_{\alpha\beta}(r) = 4\varepsilon_{\alpha\beta}[(\sigma_{\alpha\beta}/r)^{12} - (\sigma_{\alpha\beta}/r)^6] \quad ,
\end{equation}

\noindent
where the interaction parameters are $\sigma_{\alpha\beta}$ and $\varepsilon_{\alpha\beta}$ are given in Ref.~\cite{kob1994scaling}. Schr{\o}der and Dyre showed that if the $AA$ and $BB$ forces from this potential are  shifted and cut-off at 1.5$\sigma_{AA}$, while the one for the $AB$ interaction is at 2.5$\sigma_{AA}$, the system is not prone to crystallization~\cite{schroder}. In the present work we use $\sigma_{AA}$ and $\varepsilon_{AA}$, as units of length and energy, setting the Boltzmann constant $k_B$ equal to 1.0.

We have simulated this system using $N=32.400$ particles at a constant density of 1.200, corresponding to a box size of 30. The time step was 0.005 and the simulations were carried out at constant energy. In order to improve the statistics of the results we have done simulations that were significantly longer than the $\alpha$-relaxation time (for $T \leq 0.39$ the runs extended over $2\cdot 10^9$ time steps). In addition, we have carried out 4 independent simulations of the system at the various $T$s.\\
A comparison of the static and dynamical properties of the original model with those of the modified one shows, as long as crystallization is avoided, no significant difference of the two systems.
\\
\\
{\it Analysis of single-particle trajectories:} The MSD for particles $A$, referred in the text as $ \< r^2(t) \>$,  is computed as follows:

\begin{equation}
\label{eq:MSD}
\< r^2(t) \> = \frac{1}{N_A} \sum_{i=1}^{N_A} \langle ({\textbf{r}_i(t_0+t) - \textbf{r}_i(t_0)} )^2 \rangle_{t_0}
\end{equation}

\noindent
where $\textbf{r}_i(t)$ is the position of particle $i$ at time $t$ and $\langle \cdot \rangle_{t_0}$ indicates the average over all time origins $t_0$. In addition, results were averaged over the four independent simulated samples. 

The plateau height of the MSD, $r_c(T)$, is defined as the square-root of the MSD evaluated at the time at which the logarithmic derivative of the MSD has increased by 5\% after reaching its minimum value in the caging regime. 
\\
\\
The Self-Intermediate Scattering Function (SISF) for particles $A$, referred in the text as $F_s(\textbf{q}, t)$,  is computed as follows:

\begin{equation}
\label{eq:SISF}
F_s({\textbf{q}^*}, t ) = \frac{1}{N_A}   \sum_{j=1}^{N_A} \langle e^{-i {\textbf{q}^*} \cdot  [  \textbf{r}_j( t_0+t) - \textbf{r}_j(t_0) ]  } \rangle_{t_0}
\end{equation}

\noindent
where $\textbf{q}^*$ is the wavevector corresponding to the location of the main peak in the static structure factor $S(q)$, with $q=|\textbf{q}^*|=7.25$. The structural relaxation time $\tau_\alpha(T)$ is defined as the time at which the SISF reaches the value $1/e$.
\\
The height of the plateau of the SISF, $f_c(T)$, is computed as the value of the SISF at which the MSD reaches the value of $r_c^2(T)$.
\\
\\
To estimate the $T$-dependence of the stretching exponent $\beta$ of the SISF we measured the quantity $m(t;T)=-d \log F_s(q^*,t;T)/dt$, which is independent on the height of the plateau $f_c(T)$. Since at long times we expect the SISF to be compatible with a stretched-exponential behavior, we fitted $\log m(t;T)$ in the range $[0.5\tau_\alpha,5\tau_\alpha]$ with the expression $\mu(t;T) = a(T) + [\beta(T)-1] \log t$ to estimate $\beta$. In Fig.~\ref{figSM_beta}a we show the $T$-dependence of $\beta$ as obtained from this fitting procedure.
In Fig.~\ref{figSM_beta}b we plot the ratio $\epsilon(t;T) = m(t;T)/\mu_{2/3}(t,T)$, with $\mu_{2/3}=a'(T)-\frac{1}{3}\log t$ being the best fit with the expression $\mu$ to the $\log m$-dataset, if $\beta=2/3$ is kept constant. 
This representation documents the approach, as temperature is decreased towards the lowest $T$'s, to the master-curve behavior with $\beta=2/3$ of the SISF's described in the main text.
\\
\\
{\it Analytical approximation for the contribution of the tails to the total MSD:}
The contribution $ \langle r^2(t)\rangle _{\rm tail} $ mentioned in the main text is defined as: 
\begin{equation}
\begin{aligned}
    \langle r^2(t)\rangle _{\rm tail} = 3 \langle x^2(t) \rangle_{\rm tail}
     &= 3 \int_{x_0(t)}^{+\infty} x^2P(x,t) \, dx  
\end{aligned}
\end{equation}
where $x_0=X_0\sqrt{\langle x^2 \rangle}$ and $X_0=6$. Note that in the definition of the 1-D displacement function the density distribution is normalized such that $\int_{0}^{\infty} P(x, t) \, dx = 1$ (as already done in the main text). 
Making the change $x\rightarrow X$, and using the fact that for $X>X_0$ the distribution in $X$ is given by an exponential, see Fig.~\ref{fig:fig4}a,
we obtain:
\begin{eqnarray}
\langle r^2(t)\rangle _{\rm tail} & = & 3 \langle x^2(t)\rangle  \int_{X_0}^{+
\infty} X^2 A(t) \exp\left[- \frac{X-X_0}{L(t)}\right] \, dX  \\
& =&  \langle r^2\rangle  A(t) L(t) \left[ X^2_0 + 2X_0L(t) + 2L^2(t) \right] \\
& = & \langle r^2\rangle  A(t) L(t) X_0^2\left[ 1 + 2L(t)/X_0 + 2L^2(t)/X_0^2 \right] .
\end{eqnarray}
\\
Recalling that, by definition, the fraction of particles in the tail of the distribution is $\nu_{\rm tail}=A(t)L(t)$, we find
\begin{equation}
\begin{aligned}
     \langle r^2(t)\rangle _{\rm tail} = \langle r^2\rangle  \nu_{tail}(t) X^2_0\left[1+\phi(t)\right] \quad .
\end{aligned}
\end{equation}
Here, the function $\phi(t)= 2L(t)/X_0 + 2L^2(t)/X_0^2$ is small since $X_0$ is appreciable larger than $L(t)$, see Fig.~\ref{fig:fig4}.
Neglecting $\phi(t)$ and replacing $\langle r^2(t)\rangle$ by its short time value $r_c^2$ we then obtain, to leading order, $\langle r^2 \rangle_{\rm tail} \simeq r^2_c X^2_0 \nu_{tail}(t)$ for $t<\tau^*$, as reported in the main text.

\vspace{40mm}


\providecommand{\noopsort}[1]{}\providecommand{\singleletter}[1]{#1}%

\newpage
\clearpage

{\bf Supporting Information for\\
At low temperatures, glass-forming liquids relax in a simple way\\
Francesco Rusciano, Raffaele Pastore, Francesco Greco, and Walter Kob}\\[5mm]

\renewcommand{\thefigure}{S\arabic{figure}}
\setcounter{figure}{0}
\begin{figure}[ht]
    \includegraphics[width=1\linewidth]{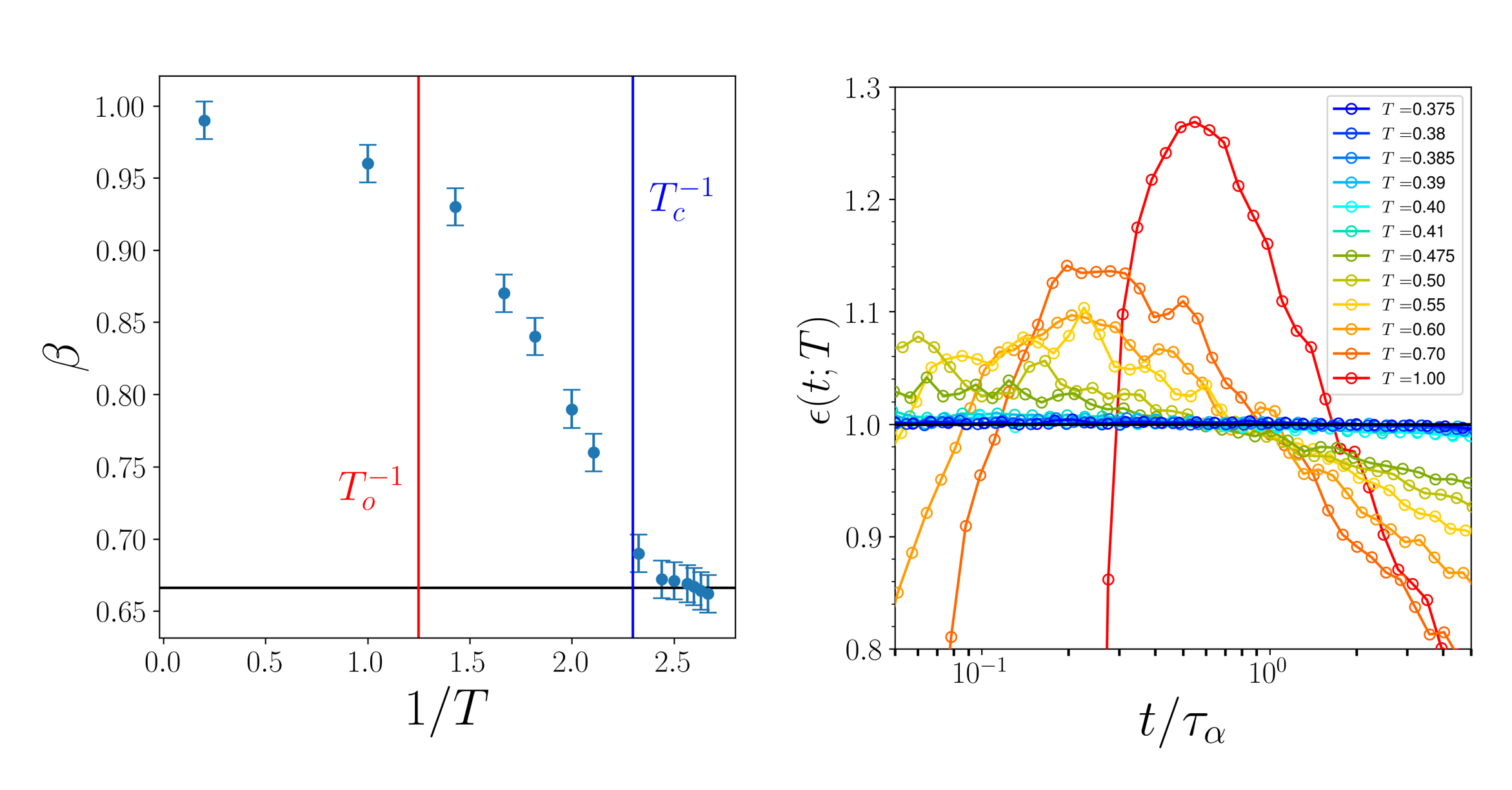}
\caption{(a) Stretching exponent $\beta$, as obtained from the intermediate scattering function, as a function of inverse temperature, at various investigated temperatures. The red vertical line indicates the critical temperature $T_c=0.435$; black horizontal line indicate the value $\beta=2/3$. (b) $\epsilon(t;T)$ as a function of rescaled time, highlighting the deviation of data at various $T$s (indicated in the legend) from the stretched exponential fit with $\beta=2/3$. 
}
\label{figSM_beta}
\end{figure}

\begin{figure}[ht]
\includegraphics[width=0.6\linewidth]{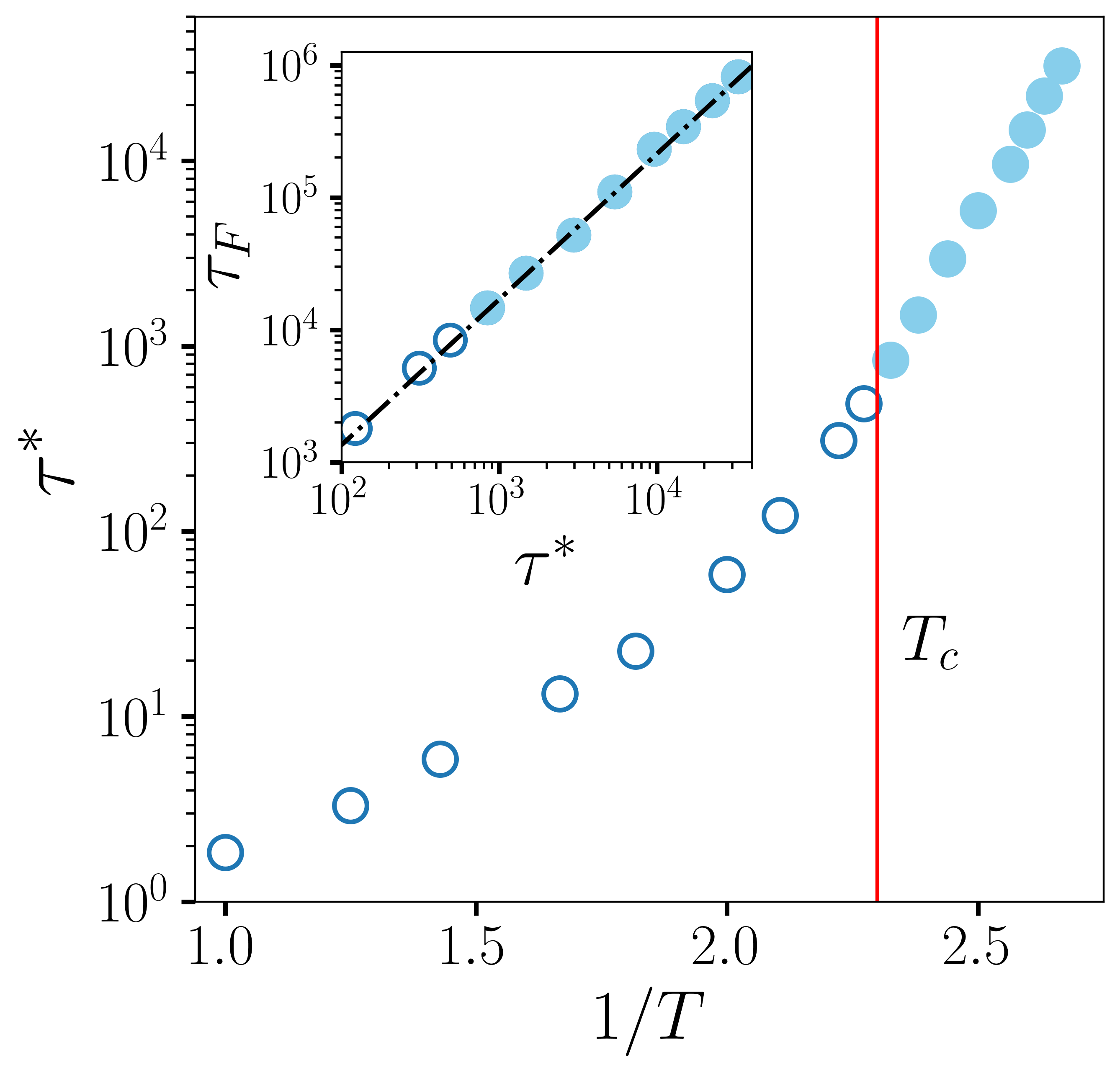}
\caption{Time of the maximum of $\alpha_2(t)$, $\tau^*$, as a function of $1/T$. The vertical line indicates the position of the critical temperature $T_c=0.435$.
Inset: Fickian time-scale $\tau_F$ vs $\tau^*$. Dashed line is a power law  $\tau_F \propto (\tau^*)^{1.1}$.
}
\label{figSM_tauF_taustar}
\end{figure}

\begin{figure}[ht]
\includegraphics[width=0.7\linewidth]{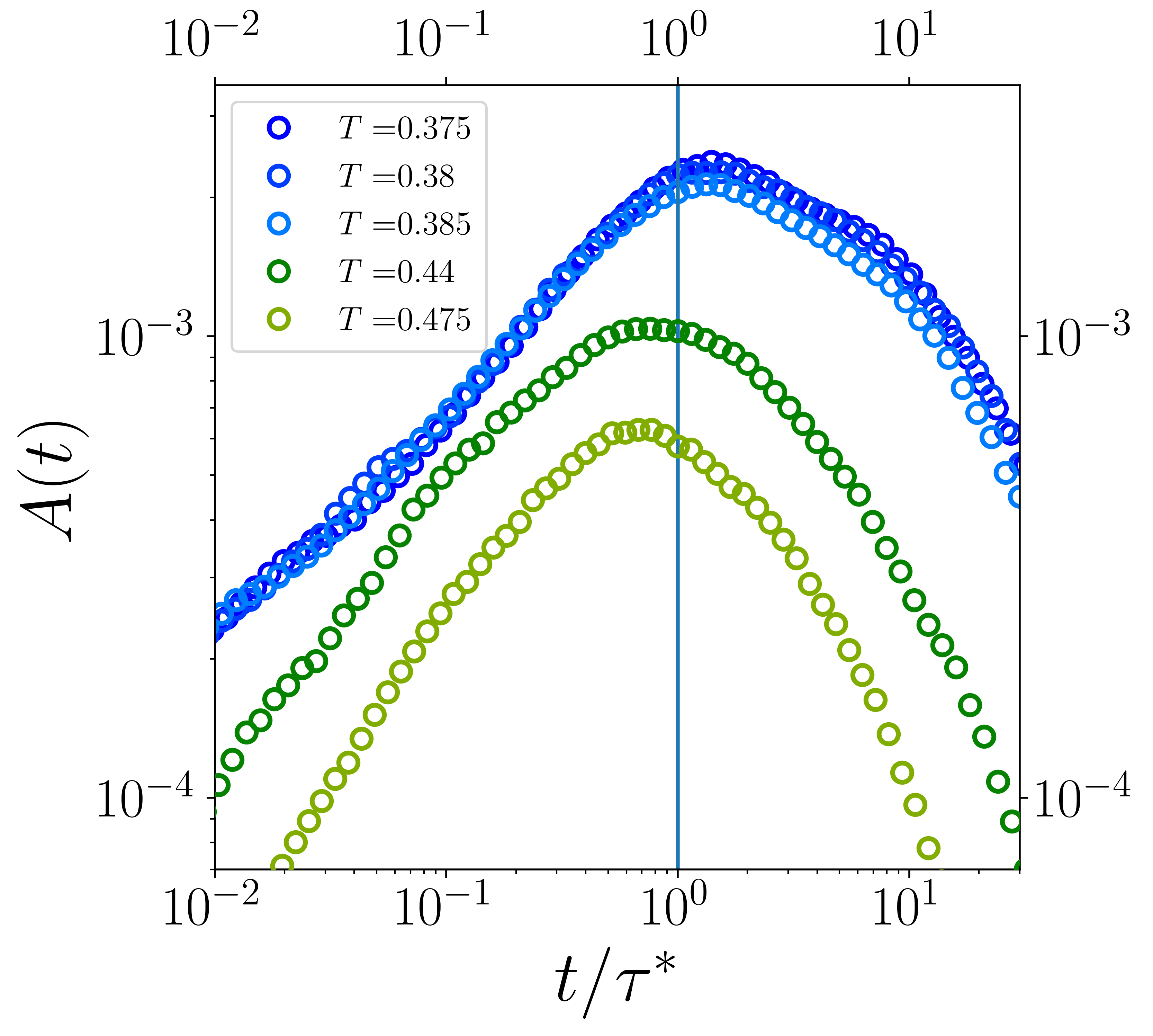}
\caption{A-dimensional pre-factor $A(t)$ of the exponential fit to the Van Hove tails as a function of rescaled time $t/\tau^*$, for temperatures above and below the critical temperature $T=0.435$, as displayed in the legend.
}
\label{figSM_A}
\end{figure}

\begin{figure}[ht]
\includegraphics[width=0.7\linewidth]{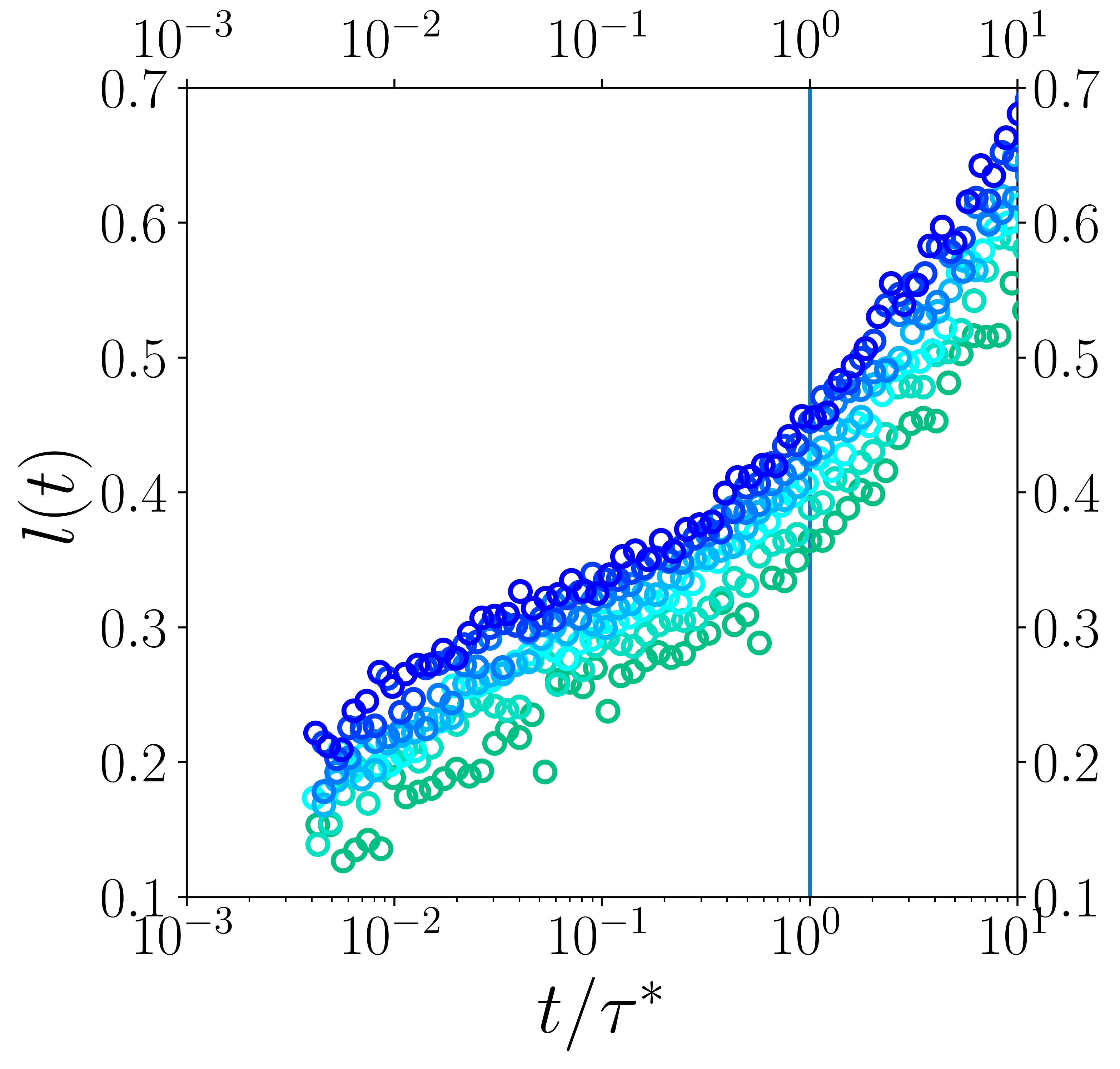}
\caption{Length scale $l(t)=L(t)\langle x^2 \rangle^{0.5}$ of the exponential fit to the Van Hove tails as a function of rescaled time $t/\tau^*$, for temperatures below the critical temperature $T=0.435$. Same colors as in the main text.
}
\label{figSM_l}
\end{figure}

\end{document}